\documentclass[12pt]{article}%
\usepackage{amsfonts, amsmath, amsthm, amsbsy}
\begin{document}
\title{Nilpotent classical mechanics: s-geometry.}  
\author{Andrzej M. Frydryszak \footnote{This work is supported by Polish KBN grant
\#1PO3B01828}\\ Institute of Theoretical Physics,\\
  University of Wroclaw,\\ pl. M. Borna 9, 50-204 Wroclaw, \\Poland}
%
%
\maketitle
\begin{abstract}
We introduce specific type of hyperbolic spaces. It is not a
general linear covariant object, but of use in constructing
nilpotent systems. In the present work necessary definitions and
relevant properties of configuration and phase spaces are
indicated. As a working example we use a D=2 isotropic harmonic
oscillator.
\end{abstract}

\maketitle

\section{Introduction}
Typical configuration space of a classical mechanical system is
endowed with the Riemannian structure if not more commonly and
simply with a flat Euclidean one. Here our aim is to study the
case of specific subclass of pseudo-Riemannian configuration
spaces, having zero trace metric. We want to restrict ourselves
here, even further, to the case of the strictly-traceless forms
which we shall call the s-forms. Definition of these notions will
be given in the next section. Motivation for this objects has
roots in the nilpotent mechanical systems cf. Ref. \cite{amf1}.
In the present work we look how such a $s$-form defined structure
influences the behaviour of mechanical system. Because lowest
possible dimension for s-geometry is equal two, we consider  the
D=2 isotropic oscillator (D=2 IHO) and compare it to the
conventional, Riemannian picture. This gives interesting
interpretation of the set of integrals of motion. For all types
of geometries defined by non-degenerate symmetric forms the
equations of motion are the same, and the set of integrals of
motion in the phase space is preserved, but the role of
particular integrals changes. In respective pictures - we switch
between $su(2)$ and $su(1,1)$ Lie algebras.

\section{s-forms}\label{sform}
Let $\mathbb V$ be a vector space $dim\mathbb{V}=2n$, over
$\mathbb R$. By $s$-form we shall understand a $\mathbb
R$-bilinear, symmetric mapping $s$
$$
s:\mathbb{V}\times \mathbb{V}\rightarrow \mathbb{R}
$$
which is weakly nondegenerate (i.e. $s(v,\, v')=0$ for all $v'
\in \mathbb{V}$, then $v=0$) and strictly traceless in the
following sense: there exists a basis $\{b_i\}_1^{2n}$ of
$\mathbb{V}$ such that
\begin{equation}
s(b_i,\, b_i)=0, \forall i
\end{equation}
Let us note here that for antisymmetric form we have automaticaly
that it is strictly traceless, moreover it is strictly traceless
for any sets of vectors.\\
The standard $s$-form is defined by
\begin{equation}\label{nat}
s(v,\, v')=\sum_{i=1}^n v_i v_{n+i}'+v_i' v_{n+i}
\end{equation}
We shall call a basis $\{e_i\}_1^{2n}$ the $s$-admissible basis
iff $s(e_i,\, e_i)=0, \forall i$. Now we are ready to define the
$s$-space. By the $s$-space we will understand a linear space
$\mathbb{V}_{2n}$ with a $s$-form and the set of all
$s$-admissible bases i.e. the triple $(\mathbb{V}_{2n}, s,
\emph{B})$. From this definition it is clearly seen that
$s$-space is not general linear covariant and its symmetry group
is essentially narrower.\\
Two $s$-spaces $(\mathbb{V},s, \emph{B})$ and $(\mathbb{V'},s',
\emph{B}')$ are isomorphic if there exists a vector space
isomorphism $\Phi$ such that
\begin{equation}
s'(\phi(v), \phi(w))=s(v, w), \quad \mbox{where}\, v,w\in
\mathbb{V}
\end{equation}
In particular we have that $\phi(\emph{B})=\emph{B}'$. Let
$Aut_s(\mathbb{V})$ be the set of automorphisms of $(\mathbb{V},
s, \emph{B}$). For the  $s$-spaces of the same dimension
respective automorphism groups are isomorphic.
\subsection{$s$-orthogonality}
The $s$-form is not positive definite and in fact is related to
the hyperbolic type of geometry. One can think about "relative
length" of vectors, in the following sense: $s(b_i,b_i)$ vanishes
but we can normalize non-vanishing product of vectors from a
fixed set, e.g. $s(b_i, b_{i+n})=1$, $i=1,2,\dots, n$. We can use
the notion of orthogonality, as usual, and structure of orthogonal
subspaces is somehow similar to the case of the symplectic
geometry \cite{ber}, but different. We will introduce relevant
subspaces in analogy to the
symplectic geometry.\\
Let $(\mathbb{V}, s, \emph{B})$ be a s-space and $W\subset
\mathbb{V}$
\begin{equation}
W^{\bot}=\{w\in\mathbb{V}|\, s(v, w)=0,\forall w\in W\}
\end{equation}
As always if $W_1\subset W_2$ then $W_2^{\bot} \subset
W_1^{\bot}$ and $(W^{\bot})^{\bot}\subset W$. Using
orthogonality, in analogy to the symplectic geometry \cite{ber},
for the s-spaces we can distinguish the following types of
subspaces:
\begin{itemize}
\item[(1)] $s$-isotropic, if $W\subset W^{\bot}$
\item[(2)] $s$-coisotropic, if $W^{\bot}\subset W$
\item[(3)] $s$-Lagrangian, if $W = W^{\bot}$
\end{itemize}
\underline{Example:}\\
$n=1$, $(\mathbb{R}^2, per, \emph{B})$, $\{b_i\}_1^2\in\emph{B}$,
where $per$ is a permanent of the matrix
$\left(
\begin{array}{ll}
v_1&w_1\\
v_2&w_2
\end{array}
\right)$.
 Then Span$\{b_1\}$
and Span$\{b_2\}$ are $s$-Lagrangian subspaces. Let us note that
in analogous example of the symplectic space $(\mathbb{R}^2, det)$
any $1$-dimensional subspace is Lagrangian.
%
%
\section{s-plectic group Ap(n)}
It is natural to ask what is the symmetry group of this sub-type
of hyperbolic geometry. The condition that trace of the s-form is
strictly $0$ (i.e. $Tr(s)|_k=0, \, k=1,\dots,2n$) restrict
essentially the group of transformations.\\
Let $\mathcal{S}=(\mathbb{R}^{2n}, s, \emph{B}\,)$ be a standard
s-space. The group of all automorphisms of $S$ preserving the
s-form (\ref{nat}) will be denoted $Ap(n)$ and called the standard
$s$-plectic group. Naturally the $Ap(n)$ is isomorphic to
$s$-plectic group $Ap(\mathbb{V},s)$ for any $2n$-dimensional
$s$-space. For further convenience we will work with matrices.
Let form $s$ be represented by the following matrix
\begin{equation}\label{natural}
s=\left(
\begin{array}{ll}
0&\mathbb{I}_n\\
\mathbb{I}_n&0
\end{array}
\right) ;\quad s^2=\mathbb{I}_{2n}; \quad s^T=s; \quad
Tr(s)|_k=0, \, k=1,2, \dots,2n
\end{equation}
The $\mathbb{I}_n$ denotes $n$-dimensional unit matrix and
$Tr(s)|_k$ the trace of a principal $k\times k$ block of the
matrix $s$. Now, $D\in Ap(n)$ when
\begin{equation}\label{inv}
D^TsD=s.
\end{equation}
In a slightly restricted way we can work in analogy to the
symplectic case. Let $s$ be given in natural basis ,
Eq.(\ref{natural}), then we can write $D\in Ap(n)$ in a block form
\begin{equation}
D=\left(
\begin{array}{cc}
P&Q\\
R&S
\end{array}
\right)
\end{equation}
with $P$, $Q$, $R$, $S$ being $n\times n$-blocks. The condition
(\ref{inv}) enforces the following relations
\begin{eqnarray}\label{cond1}
R^TP&=&-P^TR, \quad R^TQ+P^TS=\mathbb{I}_n\\
S^TQ&=&-Q^TS, \quad S^TP+Q^TR=\mathbb{I}_n \label{cond2}
\end{eqnarray}
and moreover $(det D)^2=1$ (unlike the symplectic case here the
sign of $det D$ depends on dimension of $s$-space).\\
{\underline{Example:}}\\
Let $D\in Ap(1)$,
\begin{equation}
D=\left(
\begin{array}{cc}
p&q\\
r&s
\end{array}
\right)
\end{equation}
The conditions (\ref{cond1}) and (\ref{cond2}) in this case have
two solutions
\begin{equation}
A=\left(
\begin{array}{cc}
0&q\\
q^{-1}&0
\end{array}
\right)\quad \mbox{and}\quad
 B=\left(
\begin{array}{cc}
p&0\\
0&p^{-1}
\end{array}
\right),\quad p,q\in\mathbb{R}
\end{equation}
These matrices generate $Ap(1)$, a subgroup is generated by $B$.
As it is easily seen: $det A=-1$, $det B=1$.
\section{Classical mechanics in s-spaces}\label{s-mech}
As a first application of $s$-spaces let us consider the
formalism of classical mechanics, where original configuration
space is not Euclidean or Riemannian, but is equipped with the
flat $s$-form, so the lowest possible dimension is equal two and
symmetry group is restricted to translations and $Ap(n)$
transformations. Since the form $s$ is symmetric and
non-degenerate, the formulation of conventional tools of the
classical mechanics mostly remains valid. As a fundamental system
for this type of mechanics, modeled on $\mathbb{R}^{2n}$ we shall
consider the two dimensional harmonic oscillator, for simplicity
- isotropic.
\subsection{Configuration space}
To define a mechanical system we shall introduce a Lagrangian
with terms being the analogs of kinetic and potential energy.
Analogs only, because the $s$-form is not positive definite.
\begin{equation}
L=\frac{m}{2}s(\dot{x},\dot{x})-V(x)
\end{equation}
In scope of the further example of the IHO, let
\begin{equation}
V(x)=\frac{m\omega}{2}s(x,x)=\frac{m\omega}{2}s_{ij}x^ix^j.
\end{equation}
 The form of the Euler-Lagrange equations is conventional, but due to
the new geometry, an interpretation of motions of the system gets
different.
\subsection{Phase space}
Passage to the phase space via the Legendre transformation is
straightforward and we get Hamiltonian containing $s$-form
\begin{equation}
H=\frac{1}{2m}(s^{-1})^{ij}p_ip_j+\frac{m\omega^2}{2}s_{ij}x^ix^j
\end{equation}
and conventional equations of motion
\begin{equation} \left\{
\begin{array}{rcl}
\dot{x}^i&=&\frac{1}{m}(s^{-1})^{ij}p_j \\
\dot{p}_i&=&-m\omega^2 s_{ij}x^j
\end{array}
\right.
\end{equation}
The form of  the Poisson brackets is standard, where contraction
between momenta and coordinates is, as usual, done by means of
natural pairing in cotangent bundle over the configuration space.
Using the $s$-form we can dualize coordinates, to have all phase
space indices on one level $x_i=s_{ij}x^j$. Then natural
symplectic form $\Omega$ on such a phase space will be given in
symplectic basis by the following matrix
\begin{equation}\label{bas}
\Omega=\left(
\begin{array}{cc}
0&s^{ij}\\
-s^{ij}&0
\end{array}
\right)
\end{equation}
What is obviously true, for any non-degenerate symmetric form,
like Euclidean $\delta$, pseudo-Euclidean $\eta$, as well as for
the strictly traceless  form $s$.
%
\subsection{D=2 IHO}
Using symplectic coordinates $(x_i, p_i)$, we have the following
Hamiltonian for isotropic harmonic oscillator in $D=2$
\begin{equation}
H=\frac{1}{2m}(s^{-1})^{ij}p_ip_j+\frac{m\omega^2}{2}(s^{-1})^{ij}x_ix_j
\end{equation}
and Hamilton equations of motion, whose form does not depend on
the used symmetric form: $s_{ij}$, $\delta_{ij}$, $g_{ij}$
\begin{equation} \left\{
\begin{array}{rcl}
\dot{x}_i&=&\frac{1}{m}p_i \\
\dot{p}_i&=&-m\omega^2 x_i
\end{array}
\right.
\end{equation}
Let us consider the following phase space functions %
\cite{tcvq1, tcvq2}
\begin{eqnarray}
H_0&=&\frac{1}{2m}(p_1^2+p_2^2)+m\omega^2(x_1^2+x_2^2)\\
H_1&=&\frac{1}{m}(p_1p_2)+m\omega^2x_1x_2\\
H_2&=&\frac{1}{2m}(p_2^2-p_1^2)+m\omega^2(x_2^2-x_1^2)\\
H_3&=&\omega(x_1p_2-x_2p_1)
\end{eqnarray}
They are integrals of motion not only for the Euclidean $D=2$
oscillator, but also for other geometries like hyperbolic,
including $s$-geometry. This property survives also
generalization to the curved case, to any two-dimensional spaces
of constant curvature  - the curved harmonic oscillator is
superintegrable in all such cases \cite{car}. Above functions are
related to the components of Jauch-Hill-Fradkin tensor
(\cite{jau, fra}) and saturate the following identity
\begin{equation}
H_0^2-H_1^2-H_2^2-H_3^2=0
\end{equation}
In particular we can isolate three cases which are realized by
this set of phase-space integrals of motions, related to
different geometries of the original configuration space:
\begin{itemize}
\item Euclidean - with the Hamiltonian $H_0$. The integrals of motion
$(H_1, H_2, H_3)$ generate the algebra isomorphic to the $su(2)$
and the standard symplectic form
\begin{equation}
\Omega_e= \left(
\begin{array}{rc}
0&\delta\\
-\delta&0
\end{array}
\right) ,\quad \mbox{where}\quad \delta= \left(
\begin{array}{rc}
1&0\\
0&1
\end{array}
\right)
\end{equation}
\item Hyperbolic - with the Hamiltonian $H_2$. The integrals of motion
$( H_0, H_1, H_3)$ generate the algebra isomorphic to the
$su(1,1)$ and the symplectic form in this case is the following
\begin{equation}
\Omega_h= \left(
\begin{array}{rc}
0&\eta\\
-\eta&0
\end{array}
\right),\quad \mbox{where}\quad \eta= \left(
\begin{array}{rc}
-1&0\\
0&1
\end{array}
\right)
\end{equation}
\item s-Hyperbolic - with the Hamiltonian $H_1$. The integrals of motion $(H_0,
H_2, H_3)$ generate the algebra isomorphic to the $su(1,1)$ and
the symplectic form in this case is the following
\begin{equation}
\Omega_s= \left(
\begin{array}{rc}
0&s\\
-s&0
\end{array}
\right),\quad \mbox{where}\quad s= \left(
\begin{array}{rc}
0&1\\
1&0
\end{array}
\right)
\end{equation}
\end{itemize}
The symplectic forms are given in bases analogous to that used for
$\Omega$ given by Eq.(\ref{bas}), where $\delta$, $\eta$, $s$ are
metrics defining respective geometries of the configuration
spaces.\\
\section*{Conclusions}
An interesting observation is that phase space description of the
system is not that sensitive to the specific geometry of the
configuration space, in particular question of integrability of
the system can be answered in unified way including various types
of geometry \cite{car}. In the forthcoming paper we will use the
introduced $s$-geometry to describe essentially nilpotent
mechanical systems.
%


\begin{thebibliography}{xx}
%
\bibitem{amf1} A. Frydryszak ~Czech. J. Phys. {\bf 55} (2005), 1409
%
\bibitem{ber} R. Berndt {\em "An Introduction to Symplectic
Geometry"}, Graduate Studies in Mathematics v. 26, AMS,
Providence, Rhode Island, 2001
%
\bibitem{tcvq1}G. F. Torres del Castillo, M. P. Vel\'{a}zquez Quevada  ~Rev. Mex. Fis. {\bf 50} (2006),
608
%
\bibitem{tcvq2} A. A. Mart\'{i}nez-Merino, M. Montesinos ~
{\em "Hamilton-Jacobi theory for Hamiltonian systems with
non-canonical symplectic structures"} {\tt gr-qc/0601140}
%
\bibitem{car} J. F. Cari\~{n}ena, M. F. Ra\~{n}ada, M. Santander,
T. Sanz-Gil ~J. Nonlin. Math. Phys. {\bf 12} (2005), 230
%
\bibitem{jau} J. M. Jauch, E. L. Hill ~Phys. Rev. {\bf 57} (1940),
641
%
\bibitem{fra} D. M. Fradkin ~Am. J. Phys. {\bf 33} (1965), 207






\end{thebibliography}
\end{document}